\documentclass[amsmath,amssymb,aps,prb,twocolumn,superscriptaddress]{revtex4-2}
\usepackage[utf8]{inputenc}
\usepackage[normalem]{ulem}
\usepackage[colorlinks]{hyperref}
\hypersetup{colorlinks=true,
linkcolor=blue,
filecolor=blue,
urlcolor=blue,
citecolor=blue}
\usepackage{epsfig,subfigure,float,xcolor}
\newcommand\beq{\begin{equation}}
\newcommand\eeq{\end{equation}}
\newcommand\bes{\begin{subequations}}
\newcommand\ees{\end{subequations}}
\newcommand\bea{\begin{eqnarray}}
\newcommand\eea{\end{eqnarray}}
\newcommand\non{\nonumber}
\newcommand\noi{\noindent}

\newcommand\ig{\includegraphics}

\newcommand\al{\alpha}

\newcommand\ga{\gamma}

\newcommand\De{\Delta}
\newcommand\ep{\epsilon}
\newcommand\si{\sigma}

\newcommand\Om{\Omega}

\newcommand\dg{\dagger}
\newcommand\ua{\uparrow}
\newcommand\da{\downarrow}
\newcommand\sfig{\subfigure}

\newcommand\vk{{\bf k}}

\newcommand{\mc}{\mathcal}
\newcommand{\mfr}{\mathfrak}
\newcommand{\RNum}[1]{\uppercase\expandafter{\romannumeral #1\relax}}

\begin{document}
\title{Engineering Floquet topological phases using elliptically polarized light}

\author{Ranjani Seshadri}
\email{ranjanis@post.bgu.ac.il}
\affiliation{Department of Physics, Ben-Gurion University of the Negev,
Beer-Sheva 84105, Israel}
\author{Diptiman Sen}
\email{diptiman@iisc.ac.in}
\affiliation{Centre for High Energy Physics, Indian Institute of Science,
Bengaluru 560012, India}

\date{\today}

\begin{abstract}
We study a two-dimensional topological system driven out of equilibrium by 
the application of elliptically polarized light.
In particular, we analyze the Bernevig-Hughes-Zhang model when it is 
perturbed using an elliptically polarized light of frequency $\Om$ described 
in general by a vector potential ${\bf A}(t) = (A_{0x} \cos(\Om t), A_{0y}
\cos(\Om t + \phi_0))$. (Linear and circular polarizations can be obtained 
as special cases of this general form by appropriately choosing
$A_{0x}$, $A_{0y}$, and $\phi_0$). Even for a fixed value of $\phi_0$, we 
can change the topological character of the system by changing the ratio of
the $x$ and $y$ components of the drive. We therefore
find a rich topological phase diagram as a function of $A_{0x}$, $A_{0y}$ and
$\phi_0$. In each of these phases, the topological invariant given by the
Chern number is consistent with the number of spin-polarized states present
at the edges of a nanoribbon.
\end{abstract}

\maketitle

\section{Introduction}

Topological insulators (TIs) - exotic phases of matter characterized by a gapped bulk
hosting robust, conducting boundary modes - have been the talk of the town for the last
several years. These could be three-dimensional systems with two-dimensional surface
states, or two-dimensional systems which have one-dimensional edge modes. These materials
have been studied extensively both theoretically and experimentally \cite{hasan, BHZ1,
Moore1, Moore2, fu1}. A defining feature of such systems is the existence of a
bulk-boundary correspondence, i.e., a topological invariant (for example, 
a Chern number for two-dimensional TIs) derived from the bulk bands, 
defines the properties of the boundary states.

While topological materials are, by themselves, quite interesting to study, driving them
out of equilibrium using a perturbation periodic in time constitutes a rapidly
evolving area of research
\cite{kit2010,kit2011,oka2009,gu2011,lin2011,su2012,kun2014,dora2012,thak2013,
kat2013, zhu_2014,rud2013,lin2015, car2015, xiong2016, thak2017, muk2018, zhou2018}. 
In particular, one can generate topological phases by
driving a system which was non-topological to begin with. The underlying reason for this
is that while the instantaneous Hamiltonian lies in a trivial phase, the unitary
time-evolution operator over one drive cycle is topological and has eigenstates localized
near the boundaries. Such phases are termed as Floquet topological systems, since one
employs Floquet theory - which relies on the perfect time-periodicity of the drive - in
order to analyze them.

Irradiating materials with polarized light is one of the several way of experimentally
generating such Floquet topological insulators. There have been several studies which
demonstrate that using circularly polarized light to drive materials can generate and/or
modify topological phases \cite{chen18,dut2016, per2014, mciver2020}.
However, to the best of our knowledge, there are relatively few works which have studied the
effect of the more general case of elliptically polarized light 
\cite{kit21,bayuk21,Chnafa2021,zhu_2014,fern2020}.
While the effect of using elliptically polarized light may seem to be qualitatively
similar to that of circularly polarized light in some aspects, some features are
markedly different. The deviation from circular polarization introduces an anisotropy
into the time-dependent model, thereby modifying the topological properties.

In this work we study the effect of tweaking the polarization of light and see how the
effect changes as we vary the polarization. Elliptically polarized light can be created
by superposing two linear or circularly polarized beams having a phase difference. We are
interested in studying the dependence of the topological properties of a driven system
on the phase of the polarized light as well as on the relative amplitudes in the two
directions.

We begin in Sec.~\ref{sec:BHZ} with an overview of the Bernevig-Hughes-Zhang (BHZ)
model of a two-dimensional TI and analyze the symmetries and spectrum, along with
the various phases generated by tuning the parameters of the equilibrium model.
This is followed in Sec.~\ref{sec:ell} by a brief discussion on elliptically polarised
light which is then used to drive the BHZ system in Sec.~\ref{sec:drBHZ}. The topological
properties of this driven system are found to depend on the driving parameters, namely, the $x$ and $y$ components of the oscillating 
vector potential and their
phase difference. We find multiple topological phases which are characterized
by their Chern numbers. These 
are reflected in the edge states of a nanoribbon of the BHZ system which is analyzed in
Sec.~\ref{sec:ribbon}. Finally we conclude with a summary of the results and possible future 
directions in Sec.~\ref{sec:disc}.

\section{Bernevig-Hughes-Zhang model} \label{sec:BHZ}

\begin{widetext}
\begin{figure}[htb]
\centering
\vspace{-0.5cm}
\sfig[Phase Diagram] {\ig[height=4.5cm]{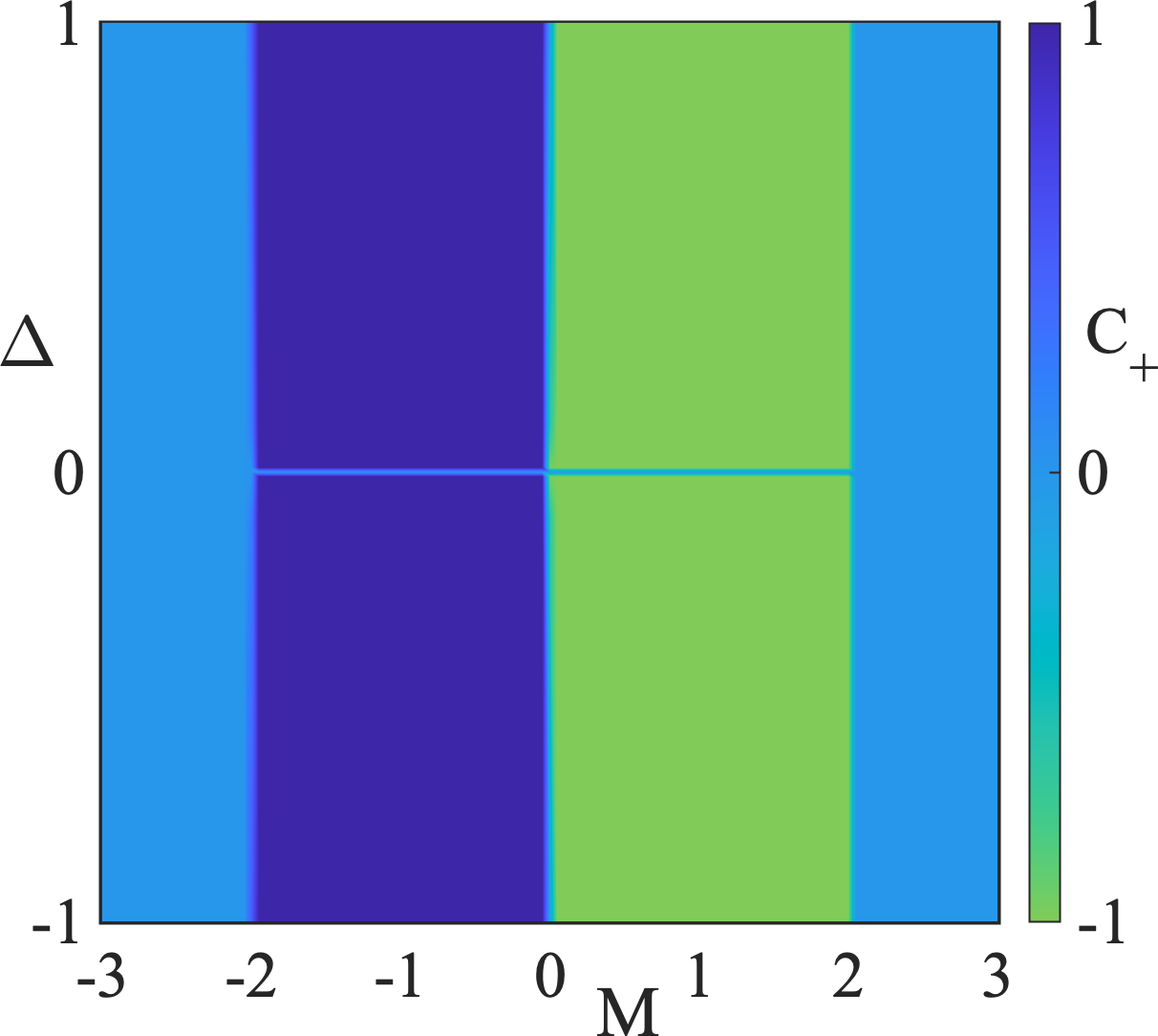}\label{fig:Eqphase}}\hspace{0.5 cm}
\sfig[Gapped Spectrum] {\ig[height=4.5cm]{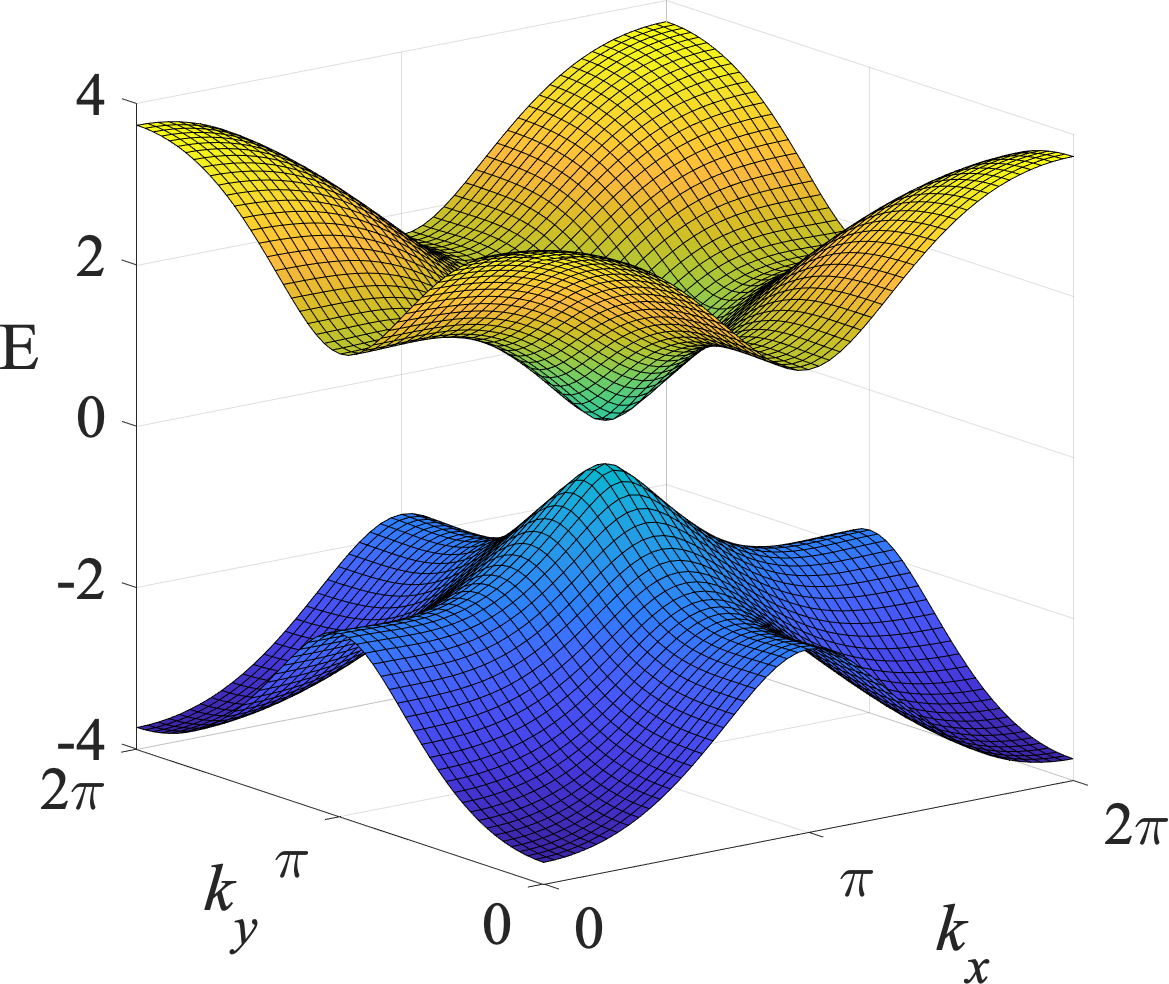}\label{fig:Eqspec}}\hspace{0.5 cm}
\sfig[Edge states] {\ig[height=4.5cm]{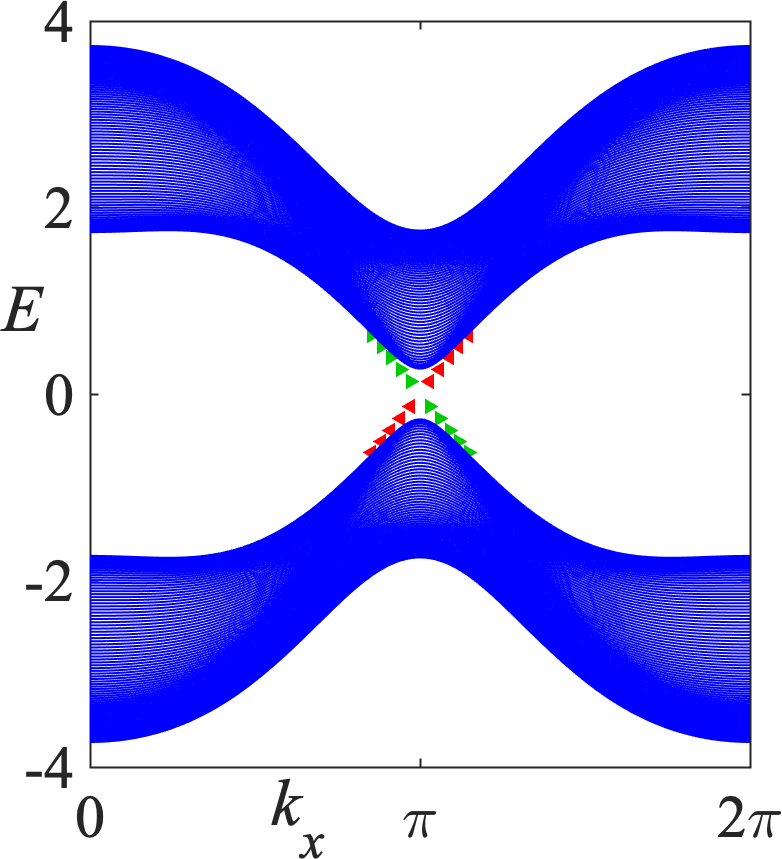}\label{fig:ribbon_eq}}
\caption{(a) Trivial and topological phases of the BHZ model. The system is in a trivial phase
when $|M|>2$. The region $-2<M<-1$ is topological with Chern number of top band $C_+ = 1$
whereas in the region $1>M>2$ has $C_+ = -1$. (b) The energy spectrum when $M = \sqrt{3}$,
$\De = \sqrt{2}$ is gapped. This corresponds to the white dot in (a) which lies in the $C_+=1$
phase. (c) Edge state spectrum for a nanoribbon parallel to the $\hat x$ direction with a finite
width $N_y = 100$ sites in the $\hat y$ direction. The bulk states are depicted in blue. 
The right (left) pointing triangles depict states with spin pointing along 
the $\pm {\hat x}$ directions, while
the color red (green) corresponds to states localized at the bottom (top) edge of the nanoribbon.}
\end{figure}
\end{widetext}

The equilibrium half-BHZ system \cite{BHZ1} with mass $M$ and spin-orbit coupling
(SOC) $\De$ is governed by the momentum-space Hamiltonian given by 
\beq
H = \sum_{\vk} \begin{pmatrix}c_{\vk,\ua}^\dag&c_{\vk,\da}^\dag\end{pmatrix}
h(\vk)
\begin{pmatrix}c_{\vk,\ua} \\ ~\\c_{\vk,\da} \end{pmatrix}
\eeq
where 
\bea
h(\vk) &=& (M + \ga \cos{k_x} + \ga \cos{k_y})\si^z\non \\ 
&+&\De(\sin{k_x}\si^x+\sin{k_y}\si^y).\label{eq:hk2by2}
\eea
Here $\si^{x,y,z}$ are the $2 \times 2$ Pauli matrices, and $\ga$ is the
hopping amplitude which we will generally set to unity
(we will also set $\hbar = 1$). This system falls under class $D$ in the
Altland-Zirnbauer classification \cite{alt97} and has the following symmetries.
\begin{enumerate}
\item \noi{\it Modified time-reversal $\mc{T}$}: While the standard time-reversal
symmetry $\Theta$ is absent, the Hamiltonian in Eq.~\eqref{eq:hk2by2}
has a modified time-reversal symmetry $\mc{T}$ which is a product of $\Theta$ and
a mirror reflection $\mc{M}_x$ about the $k_x=0$ line, i.e,
$\mc{T}h(\vk)\mc{T}^{-1} =h(\vk)$ where $\mc{T} = \mc{M}_x \Theta$.
\item \noi{\it Four-fold rotation $\mc{C}_4$}: $h(\vk)$ has a four-fold rotation symmetry
about the $z-$axis, i.e., $\mc{C}_4 h(\vk) (\mc{C}_4)^{-1}= h (\mc{C}_4 \vk)$. The operator 
$\mc{C}_4=e^{-i(\pi/4) \si^z}$ rotates the spins about the $z-$axis and transforms the
momenta as $\mc{C}_4 (k_x,k_y) =(k_y,-k_x)$.
\item \noi{\it Particle-hole symmetry ${\mc P}$}: We see from  Eq.~\eqref{eq:hk2by2}
that the system has a charge conjugation or particle hole symmetry such that
$\mc{P}h(\vk)\mc{P}^{-1} = -h^{*}(-\vk)$. This is reflected in the symmetry of the bands
about the $E=0$ plane in Fig. \ref{fig:Eqspec}.
\end{enumerate}

For a non-zero $\Delta$, the spectrum is gapped, in general, except when $M=0,\pm 2$.
We calculate the Chern number $C_{+(-)}$ for the top (bottom) band using the method
prescribed in Fukui et. al.~\cite{fukui2005}. 
The phase-diagram for this half-BHZ system at equilibrium
is shown in Fig. \ref{fig:Eqphase}. Every gap closing is accompanied by a change in the Chern
number. We see from the phase diagram that,
\bea
C_+ = \begin{cases} 
+1,& \text{~~for~~} -2<M<0, \\ 
-1,& \text{~~for~~} 0<M<2 \\
0, & \text{~~for~~} |M|>2. 
\end{cases}. 
\eea

In order to verify the bulk-boundary correspondence we consider an infinitely long nanoribbon 
having a finite width ($=N_y$ sites) in the $\hat y$ direction and running
parallel to the $x-$axis. Since the system has translation invariance along the
$\hat x$ direction, we take momentum $k_x$ as a good quantum number. However, 
the the finite
width along $\hat y$ breaks translation symmetry and therefore we take a finite
one-dimensional chain in real space parallel to $y-$axis. The Hamiltonian can therefore be
written as follows
\bes
\bea
H=h_{k_x}&+&\sum_{n_y}\frac{\ga}{2}(c_{n_y,\ua}^\dg c_{n_y+1,\ua}-c_{n,\da}^\dg c_{n_y+1,\da})\non \\
&-& \sum_{n_y} \frac{\Delta}{2} (c_{n_y,\ua}^\dg c_{n_y+1,\da} - c_{n,\da}^\dg c_{n_y+1,\ua})\label{eq:hr1}
\eea
where $h_{k_x}$ is a $2N_y \times 2N_y$ matrix containing the $k_x$-dependent terms, i.e.
\beq
h_{k_x} = \mathbb{I}_{N_y} \otimes \Big((M+\gamma \cos{k_x})\si^z + \De \sin{k_x}\si^x\Big).
\label{eq:hr2}
\eeq
\ees
\onecolumngrid
The edge modes obtained from here are shown in Fig. \ref{fig:ribbon_eq}. The continuum
formed by the bulk states is shown in blue. These are separated by an energy gap which host
the modes localized along the edges of the ribbon (green for top edge and red for the bottom
edge). These edge states are also eigenstates of $\si^x$. The right and left pointing arrows
correspond to states with $\si^x = \pm1$ respectively. Clearly, all the right moving modes 
(group velocity $v_g = \partial E/\partial k_x >0$) are localized on the bottom edge and have
$\si^x=-1$ whereas the left movers (i.e., $v_g < 0$) lie on the top edge and have $\si^x=1$.

Now that we have outlined the behavior of the equilibrium model,
we perturb the system using a time-periodic optical drive. Before we discuss the properties of the
driven BHZ model, we first recap the properties of polarized light and describe the form
of time-dependent perturbation we use.

\section{Elliptically Polarized Light} 
\label{sec:ell}

The most general form of the vector potential associated with elliptically polarized light is
\bea \bf{A}(t) &=& (A_{0x}\cos(\Om t),A_{0y}\cos(\Om t + \phi_0)), \label{eq:ell_vec} \eea
where $\phi_0$ is the phase difference between the $x$ and $y$ components.
The time-dependent electric field is therefore,
\bea
\bf{E}(t) &=& -\frac{\partial {\bf A} }{\partial t} \non \\
&=& (E_{0x} \sin(\Om t),E_{0y}\sin(\Om t + \phi_0)), \label{eq:ell_Efld}
\eea
where $E_{0x(y)} = \Om A_{0x(y)}$. The vector potential in 
Eq.~\eqref{eq:ell_vec} enters the momentum-space Hamiltonian via minimal 
coupling, $\vk \longrightarrow \vk+{\bf A}$. Therefore, the bulk Hamiltonian 
in Eq.~\eqref{eq:hk2by2} is modified as $h(\vk) \longrightarrow 
h(\vk+{\bf A})$.
We note that linear and circular polarization are special cases of Eq.~\eqref{eq:ell_Efld}. 
In the special case when $\phi_0=\pm \pi/2$, we obtain elliptically
polarized light with the axes of the ellipse aligned with the cardinal axes. 
Further, if $A_{0x} = A_{0y}$ and $\phi_0=\pm \pi/2$, we obtain 
left/right circularly polarized light.

\begin{figure}[htb]
\centering
\vspace{0.5cm}
\sfig[$\phi_0 = -\pi/3$]{\ig[width=4.2cm]{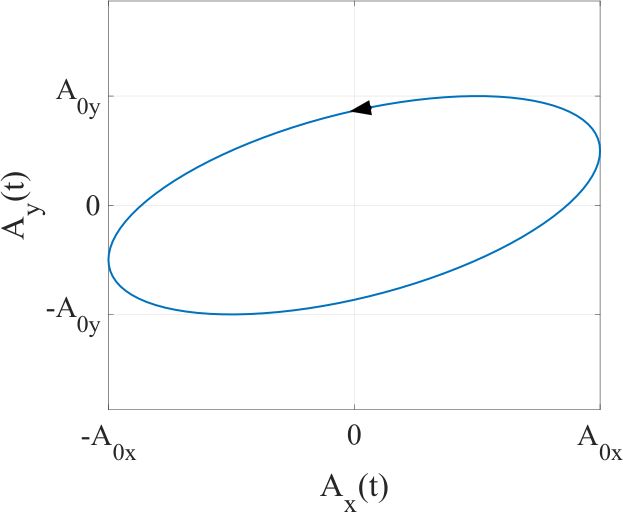}}\hspace{-0.2cm}
\sfig[$\phi_0 = -\pi/2$]{\ig[width=4.2cm]{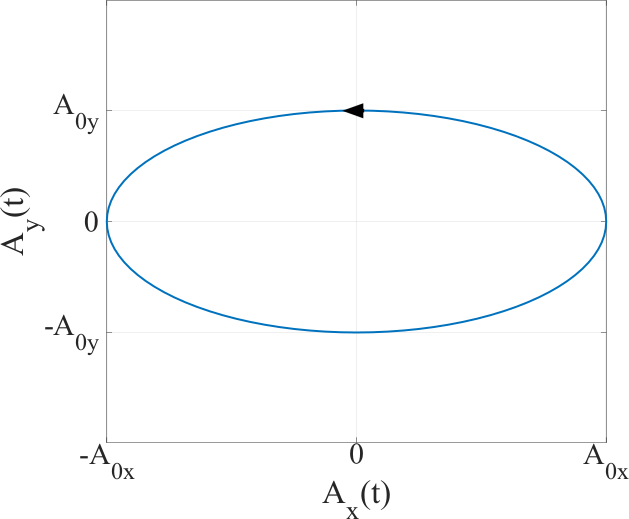}}
\caption{The polarization ellipse for two values of phase $\phi_0$.
When $\phi_0 = \pm \pi/2$, the major and minor axis of the ellipse are aligned along the 
cardinal axis. However, for other values of $\phi_0$ the ellipse is rotated. In both these
figures we have taken $A_{0x} = 0.7$ and $A_{0y}=0.4$. The ratio of these amplitudes decides
the ``flatness" of the ellipse.}
\label{fig:ell} \end{figure}

\section{Driven Topological Phases}\label{sec:drBHZ}
\begin{widetext}
\begin{figure}[h]
\centering
\begin{center}
\vspace{-1cm}\ig[width=17.8cm]{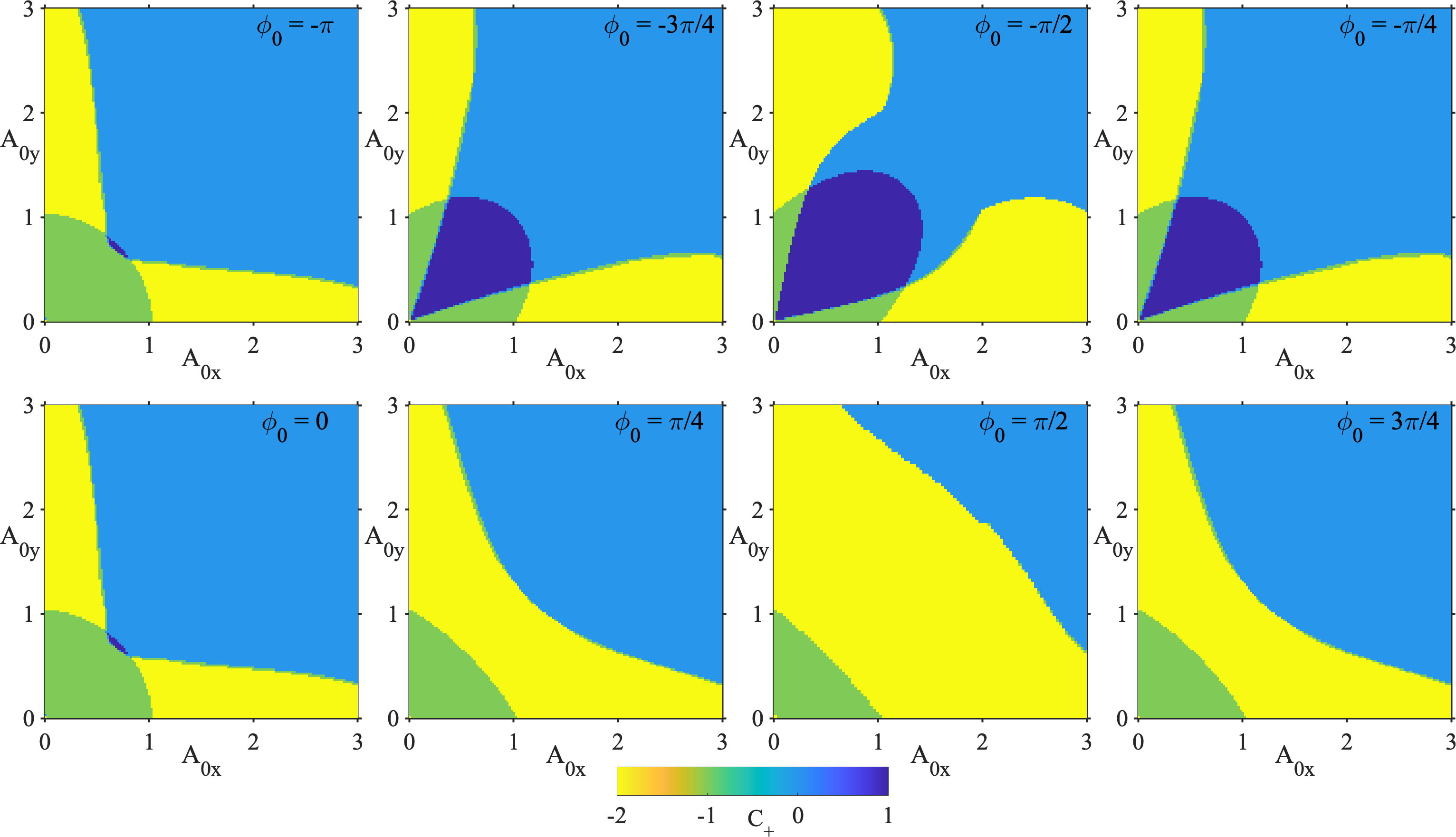}\vspace{-0.5cm}
\end{center}
\caption{Floquet Chern number $C_+$ of the positive quasienergy band, as a function
of drive amplitudes $A_{0x}$ and $A_{0y}$, for different phases $\phi_0$ of the
elliptically polarized light. Here we work in a parameter regime where the equilibrium
system is topological with $M = \sqrt{3}$ and $\Delta = \sqrt{2}$ corresponding to the
white dot marked in Fig. \ref{fig:Eqphase}. The $A_{0x} = A_{0y}$ line in (c) corresponds
to right-circularly polarized light. We have used $\Om = 5$ in all panels of this figure}
\label{fig:FloqBHZPD}
\end{figure}
\end{widetext}

We now introduce a time-dependence into the problem by using a polarized light of the form
given in Eq.~\eqref{eq:ell_vec}. The drive frequency $\Om$ is larger than the 
bandwidth
of the equilibrium system. Since the drive is assumed to be perfectly periodic, we employ
Floquet theory and calculate the quasienergy eigenvalues and eigenvectors by diagonalizing
the Floquet operator $\mc{U}_T = \mfr{T} \exp\big(-i\int_0^T dt H(t)\big)$ following the
discussion in App. \ref{sec:Floquet},
\beq
\mc{U}_T \psi_\al = e^{-i\ep_\al T} \psi_\al. \label{eq:UofT}
\eeq
The Floquet eigenvalues $\ep_\al$ are unique modulo $n\Om$ where $n$ is an integer. The
primary Floquet zone where $\ep T\in [-\pi,\pi]$ corresponds to $n=0$. The Floquet
eigenstates $\psi_\al$s are then used to calculate the Chern numbers. The results are shown as color
plots in Fig. \ref{fig:FloqBHZPD} for drive frequency $\Om = 5$. This frequency of drive is greater
than the bandwidth of the equilibrium model. Each panel corresponds to a fixed value of the phase
$\phi_0$ and shows the Chern number $C_+$ of the top band as a function of amplitudes $A_{0x}$ and
$A_{0y}$. We find that just by changing the ratio of the $x$ and $y$ amplitudes of the
elliptically polarized light, we can go from one phase to another which are topologically distinct. 
We have chosen the parameters ($M=\sqrt{3}, \De = \sqrt{2}$) such that in the absence of a drive we
are in a topological phase with Chern Number $C_+=-1$, corresponding to the white dot in Fig. 
\ref{fig:Eqphase}.

\onecolumngrid



We note two symmetries in the plots in Fig.~\ref{fig:FloqBHZPD}. First, for 
each value of $\phi_0$, the plots look the same when $A_{0x}$ and $A_{0y}$ are interchanged.
Second, the plots for $\phi_0$ and $\pi - \phi_0$ look identical. We can understand these
two symmetries as follows. Given the time-dependent periodic Hamiltonian with frequency $\Om$,
\begin{widetext}
\bea h(k_x,k_y,A_{0x},A_{0y},\phi_0,t) &=& \Big[M+\cos\big(k_x+A_{0x}\cos(\Om t)\big) + 
\cos\big(k_y+A_{0y}\cos(\Om t + \phi_0)\big)\Big]\si^z \non \\
&& +~ \Delta ~\Big[ \sin \big(k_x+A_{0x}\cos(\Om t)\big) \si^x + \sin 
\big(k_y+A_{0y}\cos(\Om t + \phi_0) \big) \si^y \Big], \label{eq:hka1} \eea
\end{widetext}
the Floquet operator is given by
\beq \mc{U}_T ~=~ {\mfr T}e^{ - i \int_0^T dt ~h(k_x,k_y,A_{0x},A_{0y},\phi_0,t) },
\label{eq:uka1} \eeq
where ${\mfr T}$ denotes time-ordering. We now observe that $\cos (\Omega t) = \cos (\Omega (T-t))$,
$\cos (\Omega t + \phi_0) = \cos (\Omega (T-t) - \phi_0)$. Also, $\si^x$ and $\si^z$ are real
whereas $\si^y$ is imaginary. These imply that an operator defined as
\bea \mc{U}'_T &=& (\mc{U}_T^{-1})^* \label{eq:up}\\
 &=& {\mfr T} e^{- i \int_0^T dt ~h'(k_x,k_y,A_{0x},A_{0y},\phi_0,t)},
\label{uka2} \eea
is the Floquet operator corresponding to a different 
time-dependent Hamiltonian given by
\begin{widetext}
\bea h'(k_x,k_y,A_{0x},A_{0y},\phi_0,t) &=& \Big[ M + \cos\big(k_x+A_{0x}\cos(\Om t)
\big) + \cos \big(k_y+A_{0y}\cos(\Om t - \phi_0)\big) \Big]\si^z \non \\
&& + ~\Delta ~\Big[ \sin \big(k_x+A_{0x}\cos(\Om t)\big) \si^x - \sin 
\big(k_y+A_{0y} \cos(\Om t - \phi_0) \big) \si^y \Big]. \label{eq:hka2} \eea
\end{widetext}
From Eqs.~\eqref{eq:UofT} and \eqref{eq:up}, we see that
\beq \mc{U}_T' \psi^*_\al = e^{-i \ep_\al T} \psi^*_\al.\label{eq:UpofT} \eeq
Hence $\mc{U}_T$ and $\mc{U}'_T$ have the same quasienergies; in particular,
the positive quasienergy band of $\mc{U}_T$ is also the positive quasienergy
band of $\mc{U}_T'$, and their eigenstates $\psi_\al$ and $\psi'_\al$ are related as 
\beq \psi'_\al ~=~ \psi^*_\al. \label{eq:psip} \eeq

Next, we see that the Hamiltonian $h'$ in Eq.~\eqref{eq:hka2} can be transformed back 
to $h$ in Eq.~\eqref{eq:hka1} in one of two ways. We can keep $\phi_0$ unchanged, 
interchange $k_x \leftrightarrow k_y$ 
and $A_{0x} \leftrightarrow A_{0y}$, shift time $t \to t + \phi_0/\Omega$ (such a shift
does not change the eigenvalues of the Floquet operator), and, finally, perform a rotation
by $\pi/2$ about the $z$-axis which transforms $\si^y \to - \si^x$ and $\si^x \to \si^y$.
(Such a rotation which is independent of $k_x, k_y$ unitarily transforms both the
Floquet operator and its eigenstates, but does not change the Chern number
defined in Eq.~\eqref{eq:chern} below).
Alternatively, we can change $\phi_0 \to \pi - \phi_0$ and $k_y \to - k_y$ but keep
$k_x$, $A_{0x}$ and $A_{0y}$ unchanged.

Finally, we consider the expression for the Chern number in, say, the positive quasienergy 
band
\bea && C_+ (A_{0x}, A_{0y}, \phi_0) \non \\
&& =~ \frac{i}{2\pi} ~\int \int dk_x dk_y \Big[ ~
\frac{\partial \psi_\al^\dg}{\partial k_x} \frac{\partial \psi_\al}{\partial k_y} -
\frac{\partial \psi_\al^\dg}{\partial k_y} \frac{\partial \psi_\al}{\partial k_x} \Big].
\label{eq:chern} \eea
We now see that the Chern number does not change if we complex conjugate $\psi$ 
(as dictated by Eq.~\eqref{eq:psip}) and either interchange $k_x \leftrightarrow k_y$, or
change $k_y \to - k_y$ but do not change $k_x$. The discussion in the previous paragraph
therefore shows that the Chern number $C_+ (A_{0x}, A_{0y}, \phi_0)$
must remain the same if we either keep $\phi_0$ unchanged and interchange $A_{0x} 
\leftrightarrow A_{0y}$, or we change $\phi_0 \to \pi - \phi_0$ but keep $A_{0x}$ and 
$A_{0y}$ unchanged. This explains the two symmetries which are visible in
Fig.~\ref{fig:FloqBHZPD}.

Now, according to the bulk-boundary correspondence, the topological character of the phase is
reflected in the presence/absence/number of edge states on a sample fashioned in the form a
ribbon, which we describe in the following section.

\section{Floquet edge Modes on a Ribbon}\label{sec:ribbon}
The Chern numbers are directly related to the number of edge states that are present
at the boundaries of a finite sample. In order to test this bulk-boundary correspondence,
we consider an infinitely long nanoribbon as we did in the equilibrium case.
When such a nanoribbon is
irradiated with polarized light described by a vector potential in 
Eq.~\eqref{eq:ell_vec}.
This is introduces a time dependence into the Hamiltonian in 
Eq.~\eqref{eq:hr2}, which we 
incorporate by minimal coupling and Peierls substitution i.e., 
\bea k_x &\rightarrow& k_x + A_{0x}\cos(\Om t), \non \\
\ga &\rightarrow& \ga e^{i A_{0y} \cos(\Om t +\phi_0)}, \non \\
{\rm and} ~~~\De &\rightarrow& \De e^{i A_{0y} \cos(\Om t +\phi_0)}. \eea

\begin{widetext}
\begin{figure}[htb]
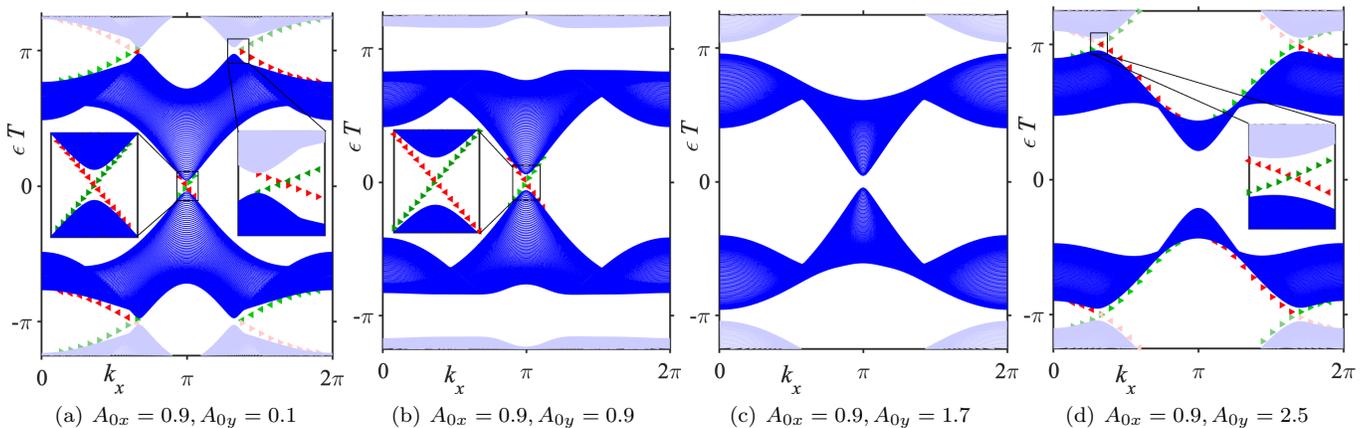

\centering
\vspace{-0.2cm}
\begin{center}
\sfig[$A_{0x} = 0.9, A_{0y} = 0.1$]{\ig[width = 4.48cm]{mpiby2_1_col.png}}\hspace{-0.1cm}
\sfig[$A_{0x} = 0.9, A_{0y} = 0.9$]{\ig[width = 4.48cm]{mpiby2_2_col.png}}\hspace{-0.1cm}
\sfig[$A_{0x} = 0.9, A_{0y} = 1.7$]{\ig[width = 4.48cm]{mpiby2_3_col.png}}\hspace{-0.1cm}
\sfig[$A_{0x} = 0.9, A_{0y} = 2.5$]{\ig[width = 4.48cm]{mpiby2_4_col.png}}
\end{center}
\vspace{-0.5cm}
\caption{Floquet edge modes on an ribbon parallel to the
$\hat x$ direction with a finite width
$N_y = 100$ sites along the $\hat y$ direction for drive frequency $\Om = 5$ with different
polarizations. These correspond to four different regions in the phase diagram shown in
Fig. 2 for $\phi_0 = -\pi/2$ with Chern numbers $C_+ = -1,1,0$ and $-2$, respectively. The
right (left) pointing triangles depict states with spin pointing along the
$\pm {\hat x}$ directions,
while the color red (green) corresponds to the bottom (top) edge of the nanoribbon. In
all the panels, the bulk states are shown in blue. The primary Floquet zone is shown in
brighter color and ranges from $\ep T = -\pi$ to $\pi$.}
\label{fig:FlEdges}
\end{figure}
\end{widetext}

We then diagonalize the Floquet operator constructed using this time-dependent Hamiltonian
to obtain the the quasienergies which are shown as a function of momentum $k_x$ in Fig.~\ref{fig:FlEdges}.
\onecolumngrid

While we have fixed $\phi_0=-\pi/2$, 
the four panels correspond to four different pairs of values of the drive
amplitudes as mentioned in the subfigure captions. All these lie in four different phases of 
Fig.~\ref{fig:FloqBHZPD}(c). 
The continuum formed by the bulk states is shown in blue with the brighter colors denoting
the primary Floquet zone (see App. \ref{sec:Floquet}) which corresponds to $n=0$, i.e,
$\ep T = -\pi$ to $\pi$. The muted colors show parts of the $n=\pm1$ Floquet zones. These Floquet bands
are separated by energy gaps which host the modes localized along the edges of the ribbon
(green for top edge and red for the bottom edge). Depending upon the ratio of $A_{0x}$ and $A_{0y}$, 
edge modes exist at $\ep T= 0 $ and/or $\ep T = \pi$. These edge states are also eigenstates of
$\si^x$. The right and left pointing arrows correspond to states with $\si^x = \pm1$ respectively.
The insets in each of the panels are zoomed-in views of the edge-state dispersion.

In Fig. \ref{fig:FlEdges}(a), we see that there are two kinds of edge states - one per edge at
$\ep T=0$ and two per edge at $\ep T = \pm \pi$. This lies in the $C_+ = -1$ phase of
Fig.~\ref{fig:FloqBHZPD}(c). On the other hand, Fig.~\ref{fig:FlEdges}(b) has only one set of
edge states at $\ep T = 0$, which is consistent with the $C_+=1$ in this phase. Fig.~\ref{fig:FlEdges}(c)
depicts $C_+=0$ phase and therefore has no edge modes,
whereas Fig. \ref{fig:FlEdges}(d) lies in the $C_+ = -2$ phase and has two
sets of edge modes, both close to $\ep T = \pm \pi$. From this we infer that a pair of edge states at
$\ep = 0$ correspond to $C_+=+1$, whereas each pair of states at $\ep T = \pm \pi$ correspond to
$C_+=-1$. These add up along with the signs to give the total Chern number $C_+$.


\section{Summary and Outlook} \label{sec:disc}

We discuss the effects of an optical drive in the form of a general elliptically polarized light on a
half-BHZ system. A range of topological phases corresponding to different Chern numbers can be generated
purely by varying the driving parameters, namely, the amplitudes of the vector potential in the $\hat x$ 
and $\hat y$ directions and their phase  difference $\phi_0$.  
We interpret this as an effect of the anisotropy that elliptically polarized light introduces into the
time-dependent Hamiltonian.

Keeping the phase $\phi_0 = -\pi/2$ and varying only the ratio of the $A_{0x}$ and $A_{0y}$ allows us 
to
tune in and out of topological phases even when we deviate away from the special case of circular polarization.
The Chern numbers are consistent with the number of spin-polarized states localized at the two edges of an
infinitely long nanoribbon with the edge states having a definite value of $\si^x$.

While the equilibrium model has phases with Chern numbers $0,\pm 1$, the time-dependent system driven
out-of-equilibrium using an elliptically polarized light allows us to generate Floquet topological phases
with higher Chern numbers as can be seen from Fig.~\ref{fig:FloqBHZPD}. Similarly, choosing the drive
parameters appropriately, the topology can even be destroyed using such an optical drive. Thus, elliptically
polarized light allows us to engineer and/or modify topological phases in the half-BHZ system.

We have confined our discussion to the case of two-dimensional topological insulators with one-dimensional
edge modes. However, the effect of elliptical polarization could have more significance in the context of higher-order
topological systems \cite{sesh19}. For instance, since generating a two-dimensional second order TI with corner
modes requires a perturbation that breaks the $C_4$ symmetry, one can expect that using elliptically polarized
light (away from the special case of circular polarization) could also achieve that \cite{ning2022}, since the
incident light (and hence the effective Floquet Hamiltonian) breaks the four-fold rotation symmetry.

\vspace*{.6cm}
\begin{acknowledgments}
R.S. thanks Devendra Singh Bhakuni and Anurag Banerjee for useful discussions.
D.S. thanks SERB, India for funding through Project No. JBR/2020/000043.
\end{acknowledgments}

\appendix

\section{Overview of Floquet Theory} 
\label{sec:Floquet}

Consider a Hamiltonian which is time-dependent and is periodic in time t, i.e.,
\beq H(t) = H(t+T), \eeq 
where $T=2\pi/\Om$, $\Om$ being the frequency.
The time-dependent Schr\"odinger equation (setting $\hbar=1$) is
\beq \Big( H(t)-i \frac{\partial}{\partial t} \Big) \Psi(t)= 0. \label{eq:TdepSch} \eeq
According to Floquet theorem \cite{floq1883, holt}, the solutions to 
\eqref{eq:TdepSch} are of the form
\beq \psi_\al(t) = e^{-i \ep_\al t} \phi_\al (t), \label{eq:ftheory} \eeq
where $\ep_\al$ is the quasienergy which is unique modulo $n\Om$, i.e.
\beq
\ep_\al \equiv \ep_\al + n\Om, ~~~~ n=0,\pm1\pm2 ...
\eeq
The state $\phi_\al(t)$ is  periodic with the same time period as the
Hamiltonian $H(t)$, i.e,
\beq \phi_\al(t) = \phi_\al(t+T). \eeq
The time evolution operator from any time $t_1$ to a later time $t_2$ is defined as
\beq \mc{U}(t_2,t_1) = \mfr{T} e^{-i \int_{t_1}^{t_2} dt H(t)}.\non \eeq
In particular, for exactly one drive cycle, this time-evolution operator is
called the Floquet operator, i.e.,
\beq \mc{U}_T = \mc{U}(T,0) = \mfr{T} e^{-i \int_{0}^T dt H(t)}.
\label{eq:FlOp} \eeq
Since $\psi(t+T)= \mc{U}_T \psi(t)$, from Eq.~\eqref{eq:ftheory},
\beq \mc{U}_T \psi_\al = e^{-i \ep_\al T} \psi_\al. \eeq
\vspace{-0.2cm}

\bibliography{refs}

\end{document}